\documentstyle[prb,aps,preprint,epsf]{revtex}

\begin{document}
\begin{center}
{\bf {\large BOUNDARY CONDITION HISTOGRAMS FOR MODULATED PHASES\ }}
\\{\today} \vspace{.5 cm}

{\it M. Benakli}\\ 
\vspace{.5cm}            
                                                                                
Department of Physics, Condensed Matter
Section, ICTP, P.O. Box 586, 34014 Trieste, Italy\\                             
   \vspace{0.5cm}                                                               
                                                                                
  {\it  M. Gabay}\\
\vspace{0.5cm}

Laboratoire de Physique des Solides
Laboratoire associ\'e au CNRS, Universit\'e de Paris-Sud,\\B\^atiment 510,
91405 Orsay Cedex, France\\
\vspace{0.5cm}

  {\it  and W.M. Saslow}\\
\vspace{0.5cm}

Department of Physics
Texas A\&M University
College Station, Texas 77843-4242\\
\end{center}
\vspace{2cm}
\noindent
{\bf PACS Number 75.10.Hk}

\vspace{1cm}
\noindent
\maketitle

\begin{abstract}
 Boundary conditions strongly affect the results of numerical
computations for finite size inhomogeneous or incommensurate structures.
We present a method which allows to deal with this problem, both for
ground state and for critical properties: it combines fluctuating
boundary conditions and specific histogram techniques. Our approach
concerns classical systems possessing a continuous symmetry  
as well as quantum systems. In particular,
current-current correlation functions, which probe large scale coherence
of the states, can be accurately evaluated. We illustrate our method on
a frustrated two dimensional $XY$ model.  

 \end{abstract}

\newpage
 Owing to their speed and power, computers are nowadays considered as
experimental tools in condensed matter physics. One of their limitations
stems from the fact that they can only model finite size systems and it is
a drawback in some cases: for instance, in the study of critical phenomena
one needs to take the infinite size limit. To overcome this problem
one may specify boundary conditions (BC). One of the most common BC
consist in imposing periodic boundary conditions (PBC). Unfortunately
PBC may lead to numerical errors and to frustration effects. 
For classical
hamiltonians on a lattice this is the case when incommensurate ground
states are favored or when interactions are long range\cite{HTD,WS}.
 Dipolar magnets
fall into that latter class: surface contributions to the
energy of the system lead to inhomogeneous structures\cite{GG,KASH}.
 For quantum
systems, current carrying states are affected by boundaries (e.g in
mesoscopic rings \cite{MESO}). The purpose of this work
is to present a method  designed to handle numerically systems  possessing
a continuous symmetry and
sensitive to boundary effects. It combines self-determined (fluctuating)
boundary conditions (FBC)\cite{SaslowCLF,O1}
 and specific histogram techniques. The latter feature
allows to study both the ground state and the critical regime of
inhomogeneous or incommensurate structures. In particular 
current-current correlation functions are obtained straightforwardly 
and the method lends itself easily
to correction to scaling analysis in the critical regime. For the sake
of simplicity, the main characteristics of the boundary
condition histogram technique are presented for classical $XY$
spins on a $D$ dimensional lattice of linear size $L$. The validity of
our approach goes beyond $XY$ systems, and it concerns the quantum
case as well. 

We begin with an analytic derivation of the
method; starting with the fluctuating boundary ensemble, we introduce
observables. This approach is the basis of a numerical study of
incommensurate phases, including the case  when the modulation
 varies with external parameters.
 However, near commensurate-incommensurate
transitions or in the case of inhomogeneous ground states, 
 large numerical errors occur.
 We then introduce a specific histogram technique (appropriate for the
fluctuating boundary ensemble)  and we show
how it can be used to probe the nature of equilibrium states and to
gain access to thermodynamic quantities such as current-current correlation
functions, including in the critical regime. Next we illustrate the
features of the boundary histogram technique on the row model
\cite{ZSG,ZSGB,KAWA}, a
frustrated anisotropic $2D$ $XY$ model: indeed, for some anisotropy range, the
state of the system evolves from an incommensurate state to an
inhomogeneous phase upon raising the temperature.

\section{Analytic approach\label{ANACH}}

 We consider two components ($XY$) spins $\vec S(i)$ on a $D$ dimensional 
lattice of linear size $L$.  
PBC  impose an $L$-periodicity to the system:
\begin{equation}
\vec S(i_1,..,i_j+L,...,i_D)=\vec S(i_1,..,i_j,...,i_D)\;;\;\forall
(i_1,..,i_j,...,i_D)  \label{PBC0}
\end{equation}

As mentioned earlier, when the spins want to form  an incommensurate
structure, PBC
generate frustration and introduce systematic numerical errors. 
Also, inasmuch as the pitch of the spiral may vary with $T$ and 
other parameters, these BC
have to smoothly evolve when one changes external parameters like the
anisotropies or the temperature\cite{HTD1}.

(Self-consistent) Fluctuating Boundary Conditions have been proposed to
overcome these problems\cite{SaslowCLF,O1}. The main feature of FBC is to add new
dynamical variables $\Delta_\alpha $ ($\alpha =1,2,...,D$ )
corresponding to a shift at the boundaries. In equilibrium the new
``boundary variables'' $\Delta _\alpha $ will fluctuate around their
\textit{most probable value} $\Delta _\alpha ^0$.

Variants of the FBC method have also been used to accelerate the approach
to the asymptotic regime, by removing some unwanted correction to
scaling\cite{O2}. This improvement applies both to ferromagnetic and
non-ferromagnetic systems.

\subsection{Fluctuating boundary conditions\label{SEC.ANA}}

The partition function of an $L^D$ system of XY
spins with PBC is:

\begin{equation}
Z_{PBC}=\int ...\int_{-\pi }^\pi \prod_id\Phi _ie^{-\beta .\left( -\frac
12\sum_{i,j}J_{ij}\cos (\Phi _i-\Phi _j)\right) }  \label{ZCLP}
\end{equation}
where the PBC are expressed by the constraint

\begin{equation}
\Phi (\vec r+n_1L\vec u_1+n_2L\vec u_2+...+n_DL\vec u_D)=\Phi (\vec r)  \label{phiclp}
\end{equation}
for any set of integers $n_1,n_2,...,n_D$. This condition can be fulfilled by putting
the lattice on a torus.

The FBC method imposes instead the following constraint 
  \begin{equation}
\Phi (\vec r+n_1L\vec u_1+n_2L\vec u_2+...+n_DL\vec u_D)=\Phi
 (\vec r)+n_1L\Delta _1+n_2L\Delta _2+...+n_DL\Delta _D
\label{phiclf}
\end{equation}
 where $\Delta _1,...,\Delta _D$ are new \textit{dynamical} degrees of
freedom, corresponding to a shift at the boundaries. Note that using 
FBC allows us to preserve translational invariance (contrary to
the free BC). Performing a change of variables 
\begin{equation}\label{SHIFT}
 \Phi
(\vec r)=\varphi (\vec r)+\vec \Delta \cdot\vec r
\end{equation}
 with $\vec \Delta =\Delta
_1\vec u_1+...+\Delta _D\vec u_D$, the constraint on $\varphi $ becomes

\begin{equation}\label{NEWPHI}
\varphi (\vec r+n_1L\vec u_1+...+n_DL\vec u_D)=\varphi (\vec r)
\end{equation}
In terms
of the new variable $\varphi $ the partition function of the $L^D$
system with FBC is:

\begin{equation}
Z_{FBC}=L^D\int_{-\pi /L}^{\pi /L}d^D\Delta \left( \int ...\int_{-\pi }^\pi
\prod_id\varphi _ie^{-\beta .\left( -\frac 12\sum_{i,j}J_{ij}\cos (\varphi
_i-\varphi _j+\vec \Delta .(\vec r_i-\vec r_j))\right) }\right)  \label{ZCLF}
\end{equation}
Wrapping the lattice around a torus automatically enforces the constraint
(Eq.\ref{NEWPHI}). It is important
to note that the integration in Eq.\ref{ZCLF} is over an interval of size
$2\pi $ for $\varphi _i$ , whereas it is over a $2\pi /L$ interval 
for $\Delta _1,...,\Delta _D$. The integration range reflects the
periodicity of the Hamiltonian.

$Z_{FBC}$ can be factorized as a product of a set of partition functions, $%
Z(\vec \Delta )$, each one corresponding to a fixed shift $\vec \Delta$ at
the boundaries.

\begin{equation}
Z_{FBC}=L^D\int_{-\pi /L}^{\pi /L}Z(\vec \Delta )d^D\Delta =L^D\int_{-\pi /L}^{\pi
/L}d^D\Delta e^{-\beta L^Df(\vec \Delta )}  \label{ZCLF2}
\end{equation}
where $f(\vec \Delta )$ is the $\frac{2\pi }L$ periodic free energy density
associated with the shift  $\vec \Delta $ at the boundary:

\begin{equation}
 f(\vec
\Delta )=-T\ln (Z(\vec \Delta ))/L^D
\end{equation}

 For $\vec \Delta =\vec
0$, we recover the PBC\ case , i.e. $Z_{PBC}=Z(\vec \Delta =\vec 0)$ and
\\
$f_{PBC}=f(\vec \Delta =\vec 0)$.

\subsubsection{The low temperature regime}

First let us consider the case of a ``ferromagnetic phase'' at low
temperature. The integral (Eq.\ref{ZCLF2}) is dominated by the minima of
$f(\vec \Delta )$; so here, $\vec \Delta =\vec 0$ . We are then allowed
to make an expansion of the integrand around $\vec \Delta =\vec 0$; for a lattice
with inversion symmetry, we get:

\begin{equation}
Z_{FBC}=L^D\int_{-\pi /L}^{+\pi /L}d^D\Delta \;e^{\left[ -\beta L^D\left\{
f_{PBC}+\frac 12\left. \frac{\delta ^2f(\vec \Delta )}{\delta \Delta _1^2}%
\right| _0\Delta _1^2+...+\frac 12\left. \frac{\delta ^2f(\vec \Delta )}{\delta
\Delta _D^2}\right| _0\Delta _D^2+O(\Delta ^4)\right\} \right] }
\label{ZCLF3}
\end{equation}
The second derivatives of the free energy are related to the components $%
\gamma^{11},...,\gamma^{DD}$ of the spin rigidity tensor
$\skew 2\bar{\bar{ \bbox{\gamma}}}$
by a geometrical
 factor $\rho $ 

\begin{equation}
\gamma^{11}=\rho \left. \frac{\delta ^2f(\vec \Delta )}{\delta \Delta
_1^2}\right| _0,...,\;\;\;\gamma^{DD}=\rho \left. \frac{\delta ^2f(\vec
\Delta )}{\delta \Delta _D^2}\right| _0  \label{gamma}
\end{equation}
(e.g, in 2D, $\rho $ is $1$ for the square lattice, and $\frac
 2{\sqrt{3}}$ for the
 triangular lattice).

We rewrite Eq.\ref{ZCLF3} , after rescaling  $\vec \Delta $, $\vec
\Delta ^{\prime }=L\vec \Delta$ :

\begin{equation}
Z_{FBC}\simeq Z_{PBC}\int_{-\pi }^{+\pi }d^D\Delta ^{\prime }\exp \left[
-\frac 1{2\rho} \beta L^{D-2}\left( \gamma^{11}
{\Delta ^{\prime }}_1^2+...+\gamma
^{DD}{\Delta ^{\prime }} _D^2\right) \right]  \label{ZCLF4}
\end{equation}
 We deduce the formal expression connecting  $Z_{FBC}$ and $Z_{PBC}$ at low
temperature, by extending the domain of integration of $\Delta ^{\prime } $ to $\left[
-\infty ,+\infty \right] $ :

\begin{equation}
Z_{FBC}=Z_{PBC}\left( \frac{(2\pi \rho)^{D\over 2} }{(\beta L^{D-2})^{D\over 2}
 \gamma }+O(\frac
1{\beta ^{{D\over 2}+1}})\right)  \label{ZCLF5}
\end{equation}
where $\gamma =\sqrt{\gamma^{11}\cdot\gamma^{22}\cdot ... \cdot\gamma^{DD}}$ .

For a system with a helical phase at low temperature, the same analysis can
be repeated, leading to the expressions:

\begin{equation}
Z_{FBC}\simeq Z(\vec \Delta ^0)\int_{-\pi }^{+\pi }d^D\Delta ^{\prime }\exp
\left[ -\frac 1{2\rho} \beta L^{D-2}\left( \gamma^{11}
{\Delta ^{\prime }} _1^2+...+\gamma
^{DD}{\Delta ^{\prime }}_D^2\right) \right]  \label{ZCLF6}
\end{equation}
and

\begin{equation}
Z_{FBC}=Z(\vec \Delta ^0)\left( \frac{(2\pi \rho)^{D\over 2} }
{(\beta L^{D-2})^{D\over 2}
\gamma }+O(\frac
1{\beta ^{{D\over 2}+1}}\right)
\label{ZCLF7}
\end{equation}

\begin{equation}
\gamma^{11}=\rho \left. \frac{\delta ^2f(\vec \Delta )}{\delta \Delta
_1^2}\right| _{\vec \Delta ^0},...,\;\;\;\gamma^{DD}=\rho \left.
\frac{\delta ^2f(\vec \Delta )}{\delta \Delta _D^2}\right| _{\vec \Delta
^0} \label{gamma2}
\end{equation}
 Here $\vec \Delta ^0$ is the minimum of $f(\vec \Delta )$ , determined
modulo $\frac{2\pi}L$ from:

\begin{equation}
\left. \frac{\delta f(\vec \Delta )}{\delta \Delta _1}\right| _{\vec \Delta
^0}=0\;\;\;,...,\;\;\;\left. \frac{\delta f(\vec \Delta )}
{\delta \Delta _D}\right|
_{\vec \Delta ^0}=0  \label{delta0}
\end{equation}
 For a spiral phase, the pitch $\vec Q_0$ is the $\frac{2\pi }L$ 
determination of  $ \vec \Delta ^0$ such that $\varphi(\vec r)\approx 0$
in equilibrium (see Eq.\ref{SHIFT}).

From Eq. \ref{ZCLF6} we deduce\cite{SaslowCLF}
 
\begin{equation}
\gamma^{11}=\frac \rho {L^D\chi _{\Delta _1}},...,\;\;\;\gamma^{DD}=\frac
\rho {L^D\chi _{\Delta _D}} 
 \label{gammadelta}
\end{equation}
where $\forall \alpha=1,..,D\;\;$ $\chi _{\Delta _{\alpha}}= \beta<(\Delta _{\alpha}-\Delta^0
_{\alpha})^2>$  is the susceptibility
for $\Delta _{\alpha}$. The validity of Eq.\ref{gammadelta}
is restricted to the low temperature regime;  more precisely, to the
region of the phase diagram  where $\forall \alpha=1,..,D\;\;$ 
$\beta \gamma^{\alpha\alpha}>>1$. 

\subsubsection{Observables:}

Using  FBC\ implies that thermodynamic quantities such as the energy 
or the spin rigidity will be different from those evaluated with
PBC. From our previous discussion, this difference shows to leading order
in size in the incommensurate phase but it is also present for
a commensurate phase (to order $O(1/L^D)$).

For the sake of simplicity we assume in this section that  $f(\vec \Delta
)$ has its minimum at $\vec \Delta =\vec 0$ . The results, however, do not
depend on the particular value of $\vec \Delta^0$ which minimizes $f$.

For a given observable $O(\Phi _j)$ the values computed with  PBC and with
FBC are:

\begin{equation}
\left\langle O\right\rangle _{PBC}=\frac 1{Z_{PBC}}\int ...\int_{-\pi }^\pi
\prod_id\Phi _iO(\Phi _j)e^{\frac \beta 2\sum_{ij}J_{ij}\cos (\Phi _i-\Phi
_j)}  \label{OCLP}
\end{equation}
\begin{equation}
\left\langle O\right\rangle _{FBC}=\frac 1{Z_{FBC}}L^D\int_{-\pi /L}^{\pi
/L}d^D\Delta \int ...\int_{-\pi }^\pi \prod_id\varphi _iO(\varphi _j+\vec
\Delta \cdot\vec r_j)e^{\frac \beta 2\sum_{ij}J_{ij}\cos (\varphi _i-\varphi
_j+\vec \Delta \cdot(\vec r_i-\vec r_j))}  \label{OCLF}
\end{equation}

The two expressions generally differ by an $O(1/L^D)$ term: for instance
the average energy density is obtained by taking the derivative of 
Eq.\ref{ZCLF7} with respect to $\beta $ yielding

\begin{equation}
e_{FBC}=e_{PBC}+\frac {D/2}{\beta L^D}+O(1/\beta ^2)  \label{ECLF}
\end{equation}
Studying  these corrections allows us to minimize
undesired finite size effects (Ref.\ref{LO2}). 

Since the FBC partition function Eq.\ref{ZCLF2} includes all 
possible BC, it is possible (at least analytically) to obtain
$\left\langle O\right\rangle _{PBC}$ directly from the FBC\ partition
function:

\begin{equation}
\left\langle O\right\rangle _{PBC}=\frac{\left\langle O\delta (\vec \Delta
-\vec 0)\right\rangle _{FBC}}{\left\langle \delta (\vec \Delta -\vec
0)\right\rangle _{FBC}}  \label{OCLP2}
\end{equation}
where $\delta (x)$ is the Dirac distribution.

For most observables, 
  $\left\langle O\right\rangle _{FBC}$ and $\left\langle
O\right\rangle _{PBC}$\ differ solely by\textit{\ corrections to scaling}.
This is, however,  not true for the spinwave stiffness.

Evaluating the spin rigidity using FBC\ is more involved: this quantity
is a (spin) current-current corelation function and measures the phase
coherence of the system. The standard
definition of $\skew 2\bar{\bar{ \bbox{\gamma}}}$, namely the response of the system to a shift 
at the
boundary, \textit{ trivially leads to a zero value}. This follows from the
very implementation of FBC, since for any imposed shift,
the dynamical variable $\vec \Delta $ can adapt itself to absorb the shift
at no cost in free energy. We  need another way to compute the
spinwave stiffness. For instance, from Eq.\ref {gammadelta} the spin rigidity   can
be computed via the susceptibility of $\vec \Delta $: one should 
keep in mind that this expression is only valid for $\beta \gamma
^{11}>>1,..., \beta \gamma^{DD}>>1$, because we have extended the domain
of integration of $\Delta ^{\prime} $ to $\left[ -\infty ,+\infty \right] $.
Near a commensurate-incommensurate boundary,  at least one of the
 $\gamma ^{\alpha\alpha}\to 0$, 
 and so one needs to use another procedure.

Another limitation of this approach concerns inhomogeneous states: in
that case one cannot use a single, constant $\vec \Delta$; for domains
separated by domain walls, for instance, each domain wall constitutes a
boundary and $\vec \Delta$ depends on spatial coordinates.
     
To overcome these problems,  we will now introduce  
$\Delta -$histograms, in conjunction with FBC.

\subsection{$\Delta -$Histograms}

In part \ref{SEC.ANA} we showed that the partition function with FBC is 
a sum over partition functions $Z(\vec \Delta )$. A practical way to perform this sum is to count the number of
 configurations obtained for each of the allowed values of $\Delta _{1},..., 
\Delta _{D}$. Since 
these quantities are defined modulo $\frac{2\pi }L$ , this can be easily done by
histograms in $\Delta _{1},...,\Delta _{D}$ which we call
$\Delta -$histograms.

From the standpoint of a Monte Carlo simulation, histograms are generated as    follows:
 we divide the range of variation of $\Delta _1,...,\Delta _D$ into
smaller sub-intervals. For each of these
we store the total number of configurations $n(\vec\Delta )$ having 
 $\Delta _1,..., \Delta _D$ within the given interval, and also the average 
of relevant quantities (such as the energy) over these configurations. 

   This yields the probability distribution $P(\vec\Delta) $
 for $\Delta$, and  averages $O^{hist}(\vec\Delta )$ of various  observables.  
  \\$P(\vec\Delta )$ is given by
\begin{equation}
P(\vec\Delta) =\frac {n(\vec\Delta )} { \mathcal{N}}
\label{proba}
\end{equation}
where $\mathcal{N}$ is the total number of generated configurations.
Similarly\\
$O^{hist}(\vec\Delta )$ is given by:

\begin{equation}
O^{hist}(\vec\Delta )=\frac 1{n(\vec\Delta )} 
\sum_{{\mathcal{C}}(\vec\Delta )}
O({\mathcal{C}}(\vec\Delta ))
\end{equation}
where $\sum_{{\mathcal{C}}(\vec\Delta )}$ is 
the sum over all 
configurations  having their boundary phase shift $\Delta_1,..., \Delta _D$     
in the same 
given histogram interval.

According to Eq.\ref{ZCLF2},  $P(\vec\Delta )$ in  
Eq.\ref{proba}
is proportional to $e^{-\beta L^Df(\vec \Delta )}$. Thus we have direct
access to the free energy dependence on $\vec \Delta $. A minimum of the
free energy translates into a maximum in $P(\vec \Delta)$, giving
both $\vec \Delta ^0$ (and thus $\vec Q_0$ the wavevector of the
incommensurate structure, see below) and $\gamma $.
 If the peak in $P(\vec \Delta)$ is
sharp enough, thermodynamic
quantities are computed for the most probable value of $\vec \Delta$;  
deviations are expected when $\gamma \to 0$ or when $P(\vec\Delta)$
 has a multi-peak
structure.

\subsection{Thermodynamics}
From our previous discussion, we can extract relevant quantities 
from a Monte Carlo study either by using FBC without histograms,
 or by using FBC with
histograms. This latter approach will give better results when the
 system is close to a phase transition. We can implement histogram
 techniques in our simulation in two ways: 

\begin{enumerate}

\item  \underline{Taking numerical derivatives of the $\Delta $-histogram
free energy:} \\ The $\Delta $-histogram free energy density is obtained from: 
\begin{equation}
f(\vec \Delta )=-\frac 1{\beta L^D}\ln \left( P(\vec\Delta )\right)
+Constant
\end{equation}
The zeroes of the first derivative of the free energy yield the value of
$\vec \Delta ^0$ . The second derivatives of the free energy computed for
$\vec \Delta=\vec \Delta ^0$ give the components of the spinwave stiffness $ \gamma $ by Eq.\ref{gamma2}. The
\textit{advantage} of this method lies in the fact that it is a fast
algorithm since \textit{no observables are computed}. Yet, 
taking the first and second numerical derivatives of the free energy
induces additional sources of errors in the results;
however this can easily be corrected by using polynomial approximations in
the vicinity of $\vec \Delta ^0$. This method is well suited for a
scaling analysis.

\item  \underline{Fluctuation-dissipation-like theorem with  $\Delta $%
-histograms:} \\ This approach is similar to the previous method, except
that the explicit expressions for the needed derivatives are computed analytically:  using 
Eqs.\ref{ZCLF}, \ref{ZCLF2} we obtain $\forall \alpha$: 
 \begin{equation}
\left. \frac{\partial f}{\partial \Delta_{\alpha} }\right| _{\Delta_
{\alpha} ^{\prime
}}=\frac 1{N}\left\langle B\delta _{\vec \Delta ^{\prime }}\right\rangle
_{FBC} \end{equation}
\begin{eqnarray}\label{FLUDIHIST}
\left. \frac{\partial ^2f}{\partial \Delta_{\alpha} ^2}\right|
 _{\Delta_{\alpha} ^{\prime
}}=\frac \rho {N}\left\{ \left\langle A\delta _{\vec \Delta ^{\prime
}}\right\rangle _{FBC}-\frac 1T\left[ \left\langle B^2\delta _{\vec \Delta
^{\prime }}\right\rangle _{FBC}-\left\langle B\delta _{\vec \Delta ^{\prime
}}\right\rangle _{FBC}^2\right] \right\} 
\end{eqnarray}
where $N$ is the total number of spins and:

\begin{equation}
A=\frac 12\sum_{i,j}\frac 12J_{ij}(\vec u_{ij}\cdot\vec 
u_{\alpha})^2\cos (\varphi
_i-\varphi _j+\vec \Delta \cdot(\vec r_i-\vec r_j))
\end{equation}

\begin{equation}
B=\frac 12\sum_{i,j}J_{ij}(\vec u_{ij}.\vec 
u_{\alpha})\sin (\varphi _i-\varphi
_j+\vec \Delta \cdot(\vec r_i-\vec r_j))
\end{equation}

\begin{equation}
\delta _{\vec \Delta ^{\prime }}=\delta \left( \vec \Delta -\vec \Delta
^{\prime }\right)
\end{equation}

This allows a better determination of the two derivatives compared to
the previous method but it implies a slowing down of the algorithm, due to
the computation of  averages\ .

\end{enumerate}

\subsubsection*{General remarks  on the $\Delta $-histogram method}

\begin{itemize}
\item   
 $\vec \Delta ^0$ measures the incommensurability of the
system; since
the partition function is $2\pi /L$
periodic, $\Delta$ is obtained  modulo $2\pi /L$. Changing $\vec \Delta$ to
$\vec \Delta +(\frac{2\pi}{L})\vec u_{\alpha}$
 corresponds to replacing the angular variables $\varphi_i$
by $\varphi_i -({2\pi\over L})\vec u_{\alpha}\cdot\vec r_i$ and leaves the
 partition function unchanged. For an  homogeneous state, the helicity\break  
 --~denoted by $\vec Q$ -- is defined modulo $2\pi$.
In the simulation we may identify its equilibrium value $\vec Q_0$ with
the value $\vec \Delta ^0$  corresponding to $\varphi\simeq 0$ 
(see Eq.\ref{SHIFT}).
 Working at low $T$ ensures
that the determination of $\vec \Delta ^0$ will not drift by increments of 
$2\pi /L$ (a variation of  $\vec \Delta ^0$ by $2\pi /L$ would entail a shift
in  $\varphi$ by $(2\pi /L)\vec u_x$, which will not occur within a
reasonable number
of Monte Carlo steps at low T). 

\item 
According to the previous discussion, at low enough $T$ one can construct 
$2\pi $ $\Delta-$histograms: the $2\pi /L$ periodicity does not show up
 in the results of the simulation. Closer to the transitions, energy
 barriers between states separated by
$2\pi /L$ become low and one needs to restrict the histogram to the relevant
$2\pi /L$ interval.

\item  Near phase boundaries, histogram techniques allow more accurate 
determinations of thermodynamic quantities. Within that framework it is
 also possible to reduce corrections to scaling, for instance by adding
 an extra term of the form $\mu (\vec\Delta -\vec\Delta ^0)^2$ to the
 partition function and thus to $P(\vec\Delta)$, and then by determining 
 $\mu $ self-consistently such that corrections to scaling be minimized
(Ref.\ref{LO2}). 

\item  Using histograms can give information about the nature of 
the commensurate-incommensurate
(C)-(IC) transition: in the incommensurate
phase, the free energy displays two minima at $\pm \vec \Delta ^0$. A
first order transition will be characterized by the coexistence of a third
local minimum in $f(\vec \Delta )$ for $\vec \Delta =\vec 0$,
 at some characteristic temperature.
\end{itemize}

\section{Application of the boundary condition histogram technique}

\begin{figure}
\begin{center}
{\parbox[t]{11.5cm}{\epsfxsize 11.5cm
\epsffile{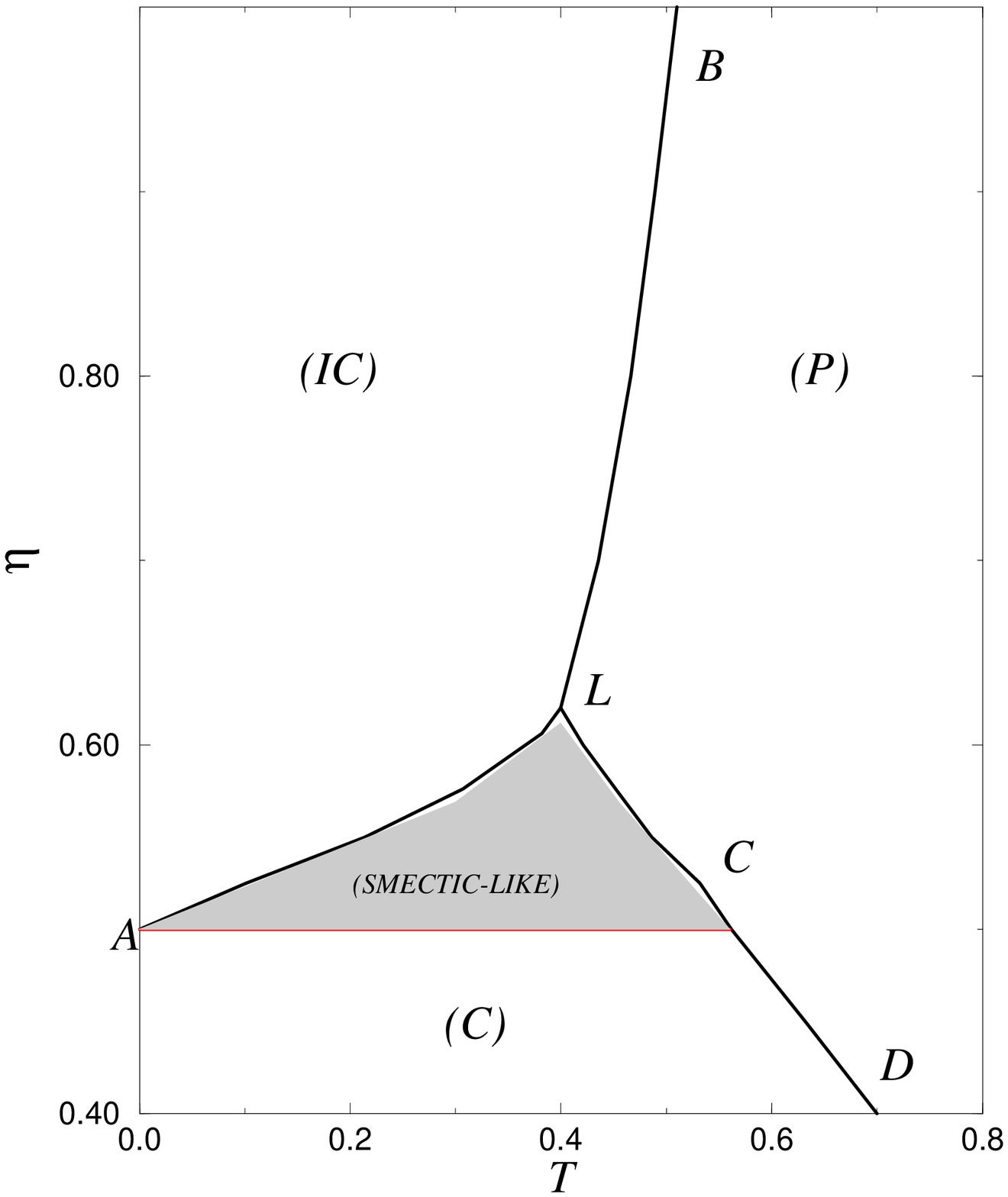}}
}
\end{center}
\hspace{-1.0truecm}
\protect\caption{
MC phase diagram of the row model, in the ($\eta, T$) plane.
}
\label{PHASEDIAG}
\end{figure}

In order to illustrate the main features of the method introduced in
section \ref{ANACH}, we consider the row model, a frustrated  anisotropic
2D XY model defined on the triangular lattice\cite{ZSG,ZSGB,KAWA}. Only
nearest
neighbor
sites are coupled: $J_{ij}=-\eta J$ $(J>0)$ for $i$ and $j$ along the
horizontal direction and $J_{ij}=-J$ otherwise. 
The case $\eta =1$ corresponds to the fully frustrated model\cite{BZG}
 but here we
will assume  $\eta \neq 1$ .
At $T=0$ the following phase diagram is obtained. For $\eta<0.5$ one has
a commensurate collinear antiferromagnetic phase. For $\eta>0.5$ an
incommensurate spiral state is observed in the horizontal 
direction denoted by $x$. The entire phase diagram of  
 the row model, in the ($\eta$, $T$) plane, was previously studied in Monte
Carlo. 
Since the incommensurability is  only present in the $x$ direction 
hybrid boundary conditions were used: PBC in the $y$ direction and FBC
{\it without} $\Delta-$histograms in the $x$ direction. Three phases
were identified (see Fig (\ref{PHASEDIAG})): a low $T$ commensurate
state (C) bounded by lines ALCD, a low $T$ incommensurate state (IC)
bounded by lines ALB and a high $T$ paramagnetic state (P). AL is
a commensurate-incommensurate transition line: starting
from $T=0$ and increasing $T$ at fixed $\eta$, $\gamma^{xx}$ 
-- the $x$ component of the spinwave stiffness
-- {\it computed via Eq.\ref{gammadelta}},    
decreases, vanishes continuously upon reaching line AL,
 increases again
 and eventually
goes back to zero on the LCD boundary. For this particular thermodynamic
path, $Q_x$, the  (incommensurate) wave vector in the $x$ direction
decreases and vanishes continuously on AL.

\noindent
The choice of this model stems from the fact that {\it (i)} as
$T$ approaches $T_{AL}$ (the temperature of the (C)-(IC) transition),
$\gamma^{xx} \to 0$ and thus Eq.\ref{gammadelta} ceases to be valid {\it (ii)}
for $T>T_{AL}$ the state of the system consists of a multidomain structure
and is not a homogeneous (C) phase (Ref(\ref{preprint}) and see below): (FBC),
 in the original
 form\cite{SaslowCLF,O1}, 
select $\vec \Delta^0$, the most probable value of $\vec\Delta$,
whereas the inhomogeneous character of the state requires more than one $\vec
\Delta$. $\Delta -$histograms then prove a useful tool to study
 the thermodynamics of the row
model in the vicinity of the (C)-(IC) transition.

In the present simulation, we used $\Delta -$histograms.
 A standard Metropolis algorithm
was applied to the spin angles and to the boundary shift in the $x$
direction. We worked with a  $48^2$ lattice and the number of MCS/spin was
 of order $10^5 - 10^6$. Typically
the first $10^4$ steps were discarded for equilibration.
We monitored $Q_x$, $\gamma^{xx}$, $\gamma^{yy}$ (the $y$ component of
the spinwave stiffness) and the staggered chiralities\cite{BZG}.
We chose $\eta=0.55$ and varied the temperature. A detailed analysis of
the simulation is presented in (Ref(\ref{preprint})). 

\begin{figure}
\begin{center}
{\parbox[t]{12.5cm}{\epsfxsize 12.5cm
\epsffile{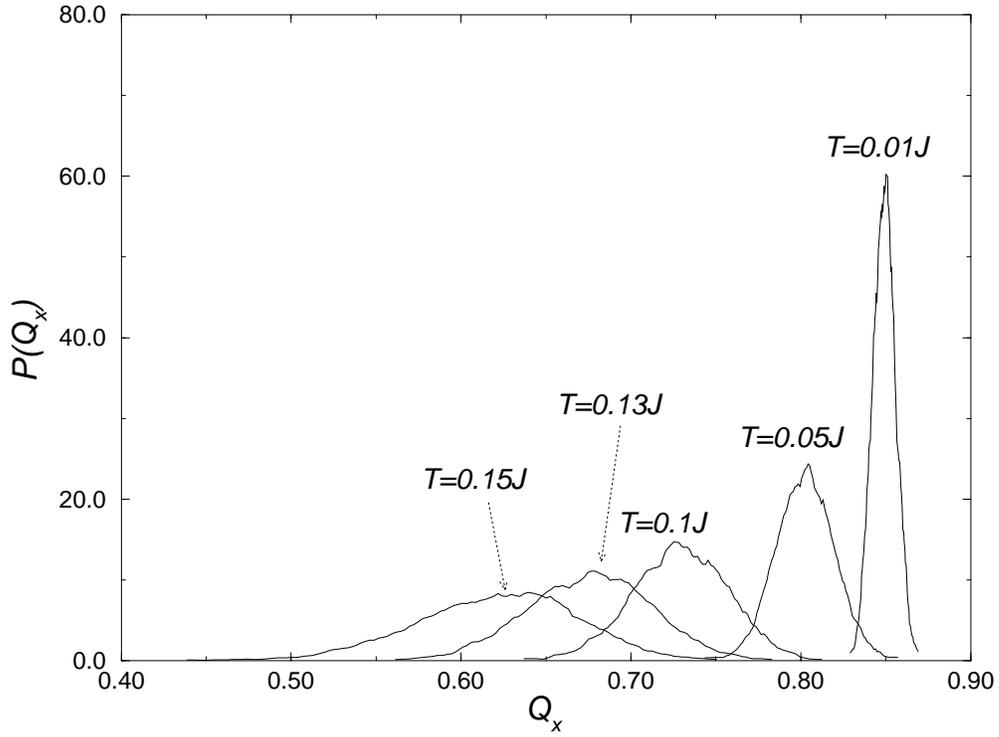}}
}
\end{center}
\protect\caption{
Unnormalized $2\pi$ $\Delta -$histogram $P(Q_x)$ for the wavevector
$Q_x$ at various temperatures. Only positive values of $Q_x$ are shown,
since $P(Q_x) = P(-Q_x)$.
}
\label{LOWTHIST}
 \end{figure}

\begin{figure}
\begin{center}
\vspace{5cm}
{\parbox[t]{12.5cm}{\epsfxsize 12.5cm
\epsffile{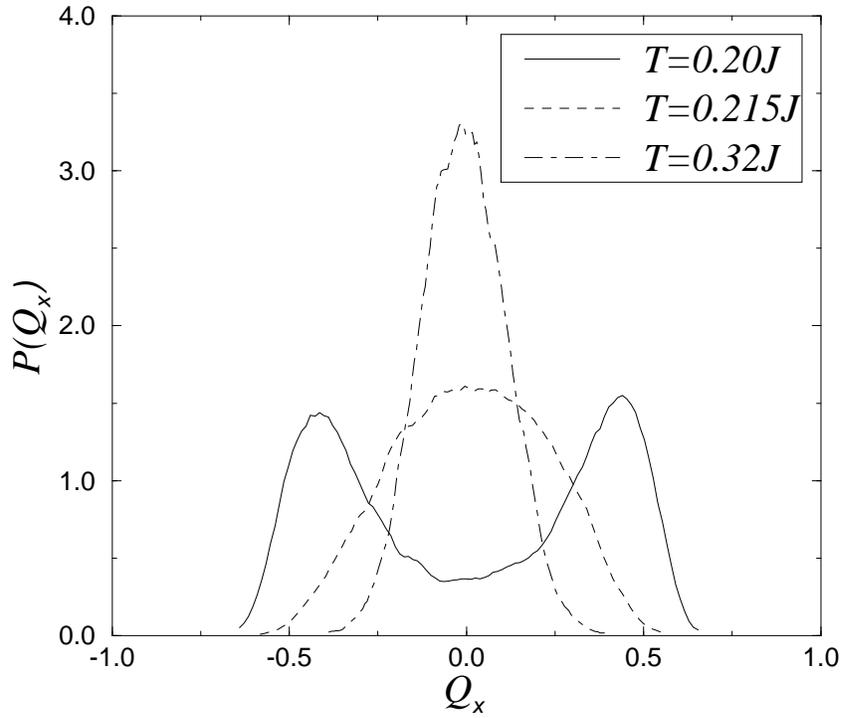}}
}
\end{center}
\vspace{-2cm}
\protect\caption{
Unnormalized $2\pi$ $\Delta -$histogram $P(Q_x)$ for the wavevector
$Q_x$ at various temperatures. For $T\geq T_{AL}\approx 0.19J$, $P(Q_x)$
has a single maximum at $Q_x=0$
}
\label{HIT2PIHIST}
 \end{figure}

\begin{figure}
\begin{center}
{\parbox[t]{13.5cm}{\epsfxsize 13.5cm
\epsffile{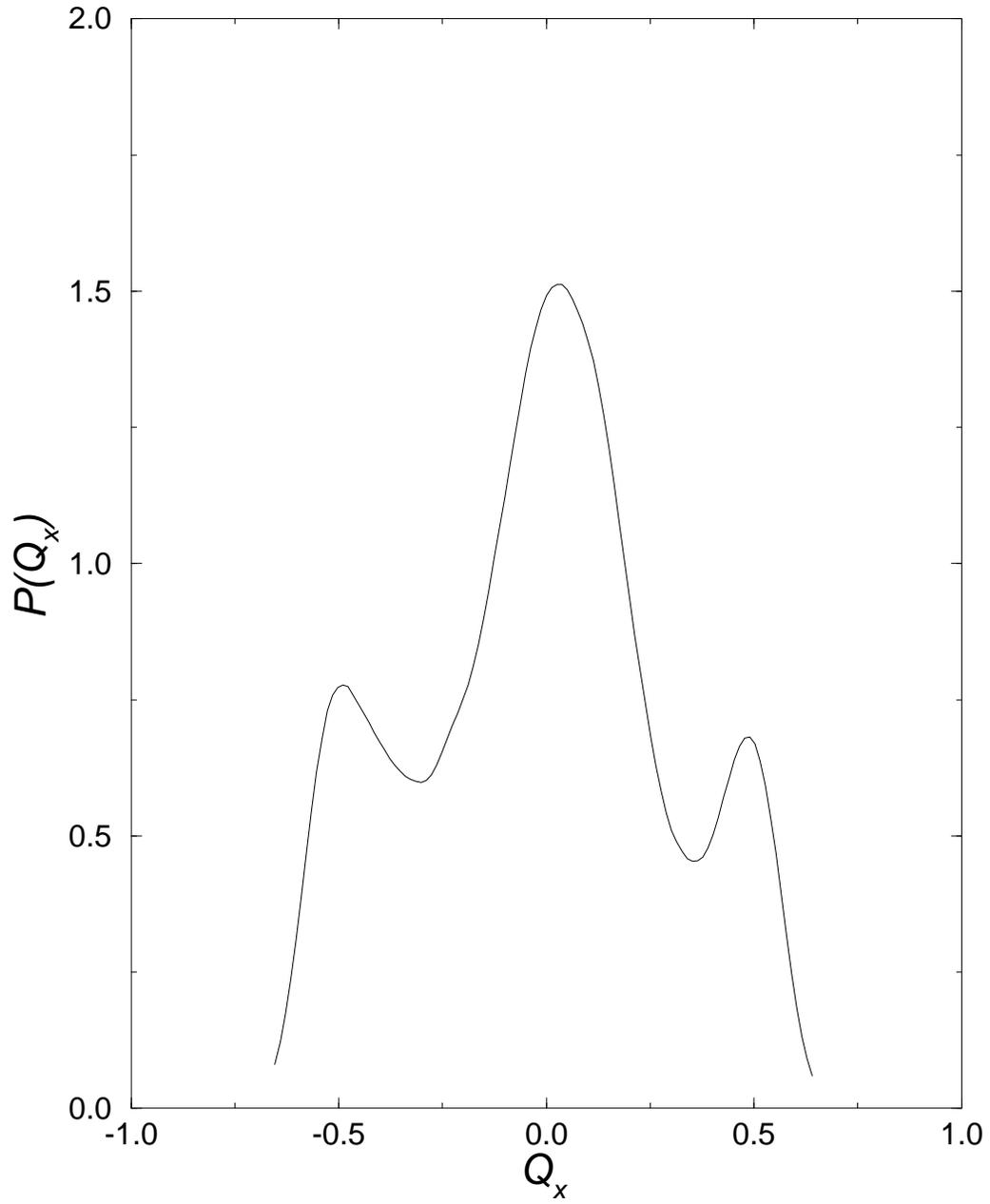}}
}
\end{center}
\protect\caption{
Unnormalized $\frac{2\pi}L$ $\Delta -$histogram  $P(Q_x)$ versus $Q_x$
 for $\eta=0.55$ and $T=0.19J$.
}
\label{HITHIST}
\end{figure}

\begin{figure}
\begin{center}
{\parbox[t]{13.5cm}{\epsfxsize 13.5cm
\epsffile{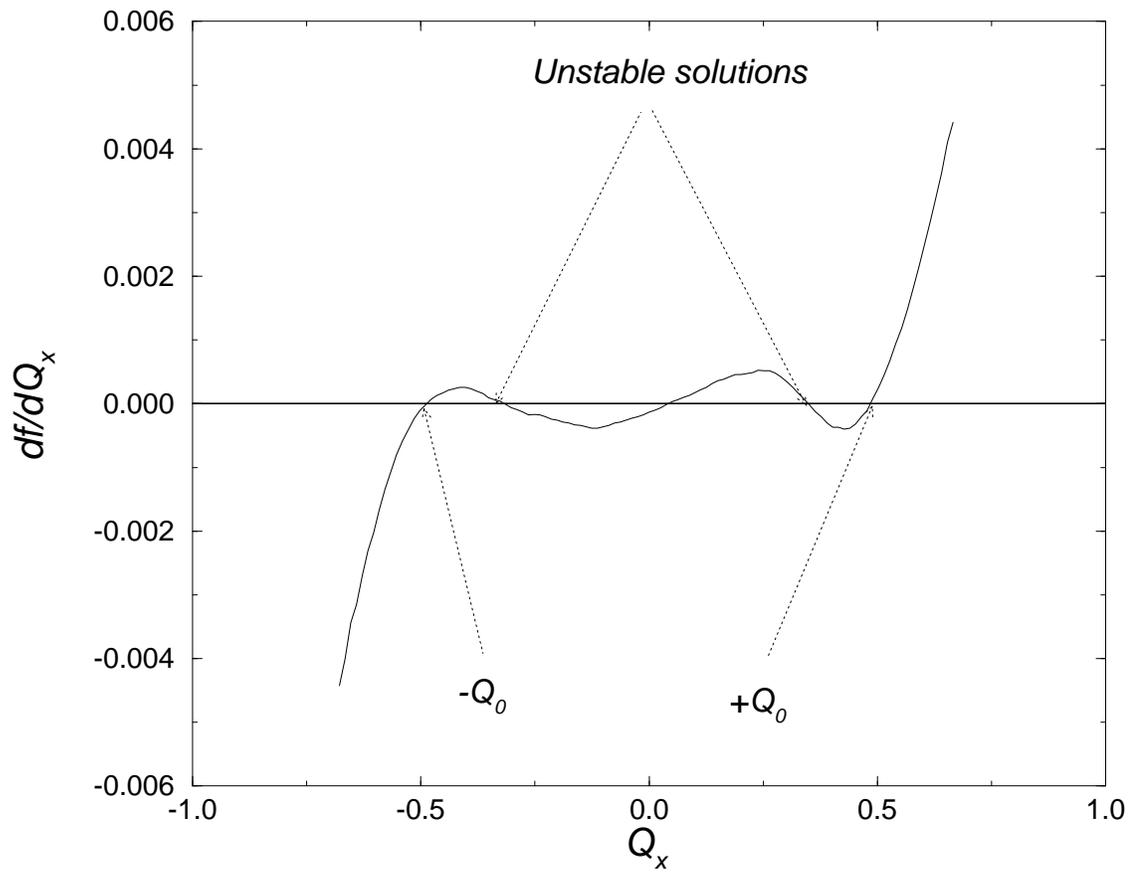}}
}
\end{center}
\vspace{2cm}
\protect\caption{
First derivative of the free energy density versus $Q_x$ for
 $\eta=0.55$ and $T=0.19J$.
}
\label{DFDQ}
\end{figure}

\begin{figure}
\begin{center}
{\parbox[t]{12.5cm}{\epsfxsize 12.5cm
\epsffile{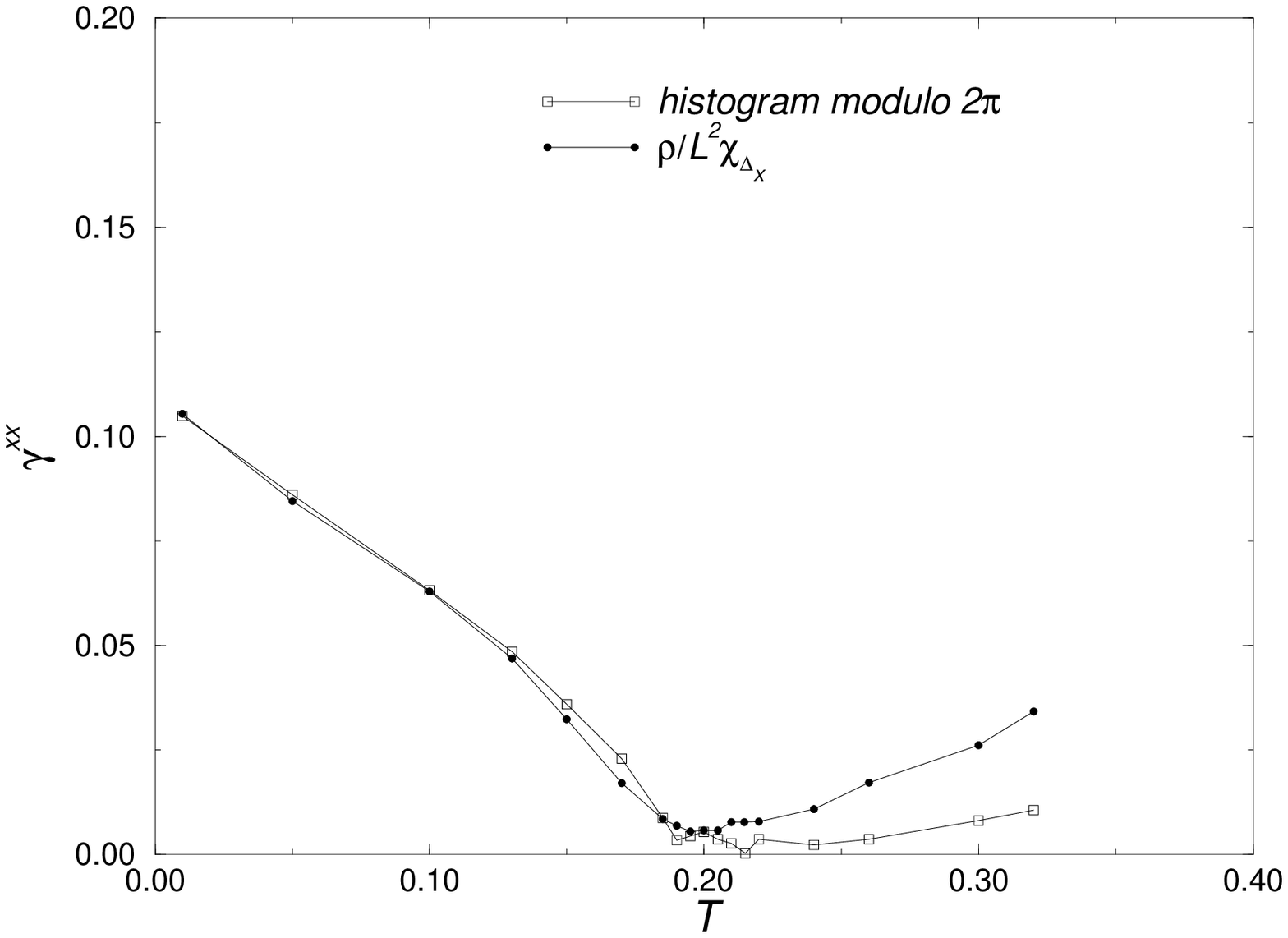}}
}
\end{center}
\vspace{2cm}
\protect\caption{
Comparison of MC data for $\gamma^{xx}$ using $\Delta$-histograms modulo $2\pi$
or using the fluctuation-dissipation theorem (Eq.\ref{gammadelta}).
}
\label{GAMHISTCHI}
 \end{figure}

\begin{figure}
\begin{center}
{\parbox[t]{13.5cm}{\epsfxsize 13.5cm
\epsffile{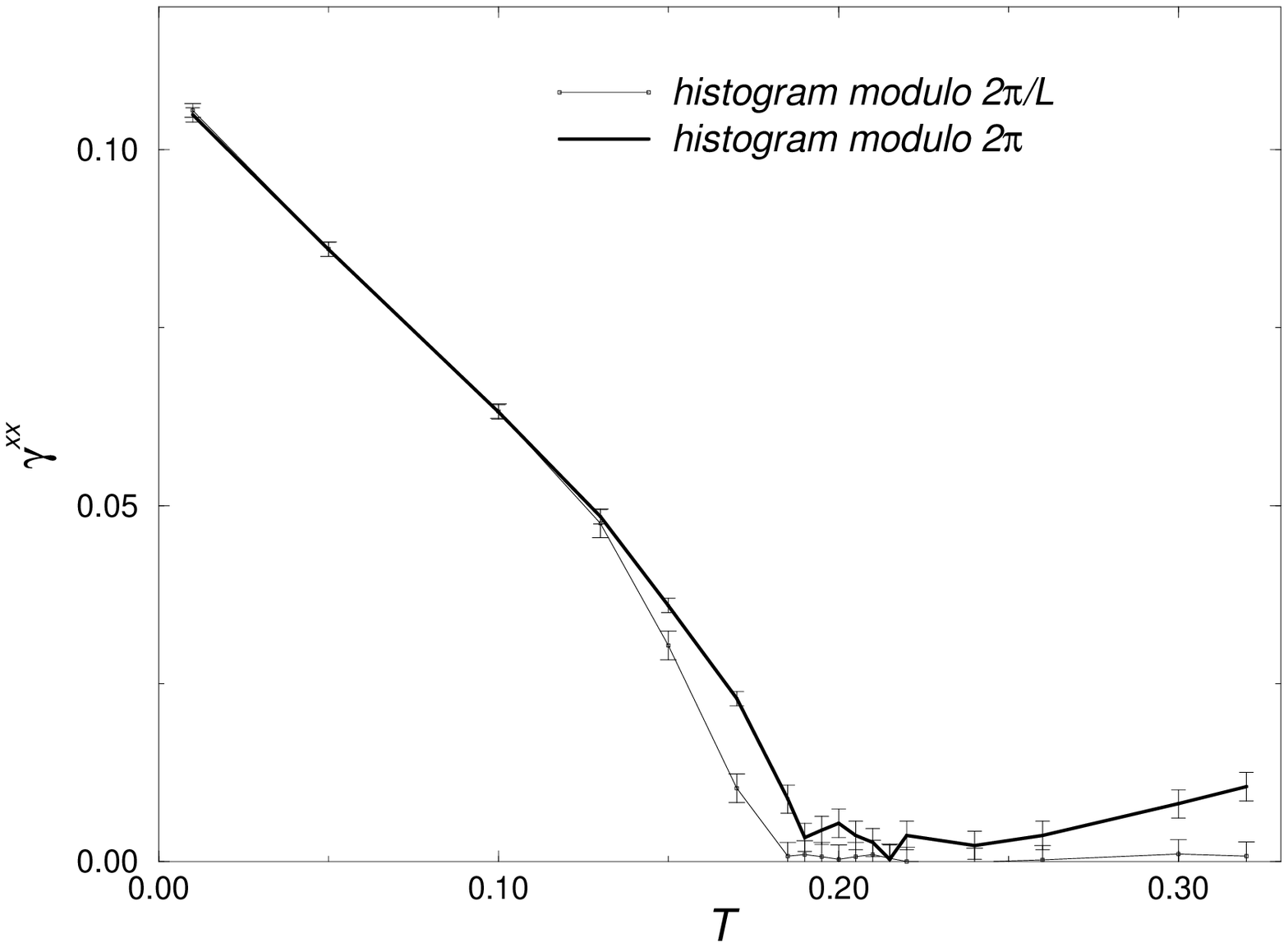}}
}
\end{center}
\vspace{2cm}
\protect\caption{
Comparison of MC data for $\gamma^{xx}$ using $\Delta$-histograms modulo $2\pi$
or modulo $2\pi\over L$.
}
\label{GAMHISTHIST}
\end{figure}

\begin{enumerate}
\item
  At low temperature, in the incommensurate phase (IC) (see  Fig
(\ref{PHASEDIAG})), we can construct $2\pi $ $\Delta -$histograms as
discussed in the general remarks of section \ref{ANACH}. The
corresponding histogram for $Q_x$ is shown in Fig (\ref{LOWTHIST}). 
At a given $T$, $P(Q_x)$ displays two well defined peaks for $Q_x=+Q_0(T)$
and $Q_x=-Q_0(T)$ (the figure only shows positive values of $Q_x$).
As the temperature brings the system closer to line AL, the peaks
at $Q_x=+Q_0(T)$ and $Q_x=-Q_0(T)$ broaden and tend to merge. In that
regime, $\vec \Delta^0$ drifts easily and if one still uses  $2\pi
$ histograms one finds that $Q_x=0$ for $T\geq T_{AL}$ (Fig
(\ref{HIT2PIHIST})). On the other hand if the $\frac {2\pi}L$
periodicity of $\Delta_x$ is enforced, one finds that $P(Q_x)$ shows
three peaks as a function of $Q_x$,  for $T\geq T_{AL}$ (Figs
(\ref{HITHIST}) and (\ref{DFDQ})). The relative weight of the
 side peaks compared to
that of the central peak ($Q_x=0$) is roughly one: 25\% of the system is
in a spiral state with wavevector $Q_x=+Q_0(T)$, 25\% is in a spiral
state with wavevector $Q_x=-Q_0(T)$ and 50\% in a collinear state. The
multi-peak structure evokes the hysteresis often present in first order
transitions but in fact it is due to the existence of an inhomogeneous
structure for $T\geq T_{AL}$ (see below).

\item
At low temperature, in the incommensurate phase (IC), 
the determinations of $\gamma^{xx}$ via Eq.\ref{gammadelta}, via $2\pi$
$\Delta -$histograms or via $\frac {2\pi}L$ $\Delta -$histograms all
agree (Figs (\ref{GAMHISTCHI}) and (\ref{GAMHISTHIST})). However 
for $T\geq 0.10J$ the three curves  move apart. In particular, for
$T\geq T_{AL}\approx 0.19J$, $\gamma^{xx}$ as obtained from $\frac {2\pi}L$
$\Delta -$histograms is zero within our error bars, whereas it is
positive with the other methods.  

$\gamma^{yy}$ does not show any singularity : it decreases as $T$
increases and vanishes on the paramagnetic boundary (line LCD of Fig
(\ref{PHASEDIAG})). 

\end{enumerate}

For $T\to T_{AL}$ from above, spins undergo critical fluctuations. At a given
$T$ the MC process generates fluctuations in $\vec \Delta^0$ by large
 increments of $2\pi/L$. If we use $2\pi$ $\Delta -$histograms
 instead of $2\pi/L$ 
$\Delta -$histograms (as one should) these fluctuations smear out the structure
of the $2\pi/L$ $\Delta -$histogram (Fig(\ref{HITHIST})). The 
$2\pi$ $\Delta -$histogram (Fig(\ref{HIT2PIHIST})) yields a single peak at
$Q_x=0$: this value is the mean of $Q_x$ averaged over the $2\pi/L$ 
$\Delta -$histogram; similarly the width of the $2\pi$ $\Delta -$histogram 
is of the order
of the average of $Q_x^2$ over the $2\pi/L$ $\Delta -$histogram. 

The multi-peak structure of the $2\pi/L$ $\Delta -$histogram also
signals the breakdown of the fluctuation-dissipation theorem:
 computing $\gamma^{xx}$ for the most probable value of $Q_x$ (i.e  $Q_x=0$)
gives a larger result than that obtained via Eq.\ref{FLUDIHIST}. The former
value of $\gamma^{xx}$ is comparable to what one gets via $2\pi$ 
$\Delta -$histograms. This explains  Fig(\ref{GAMHISTHIST}). It also explains
why $\gamma^{xx}$ determined via the most probable value of $Q_x$ is non zero 
 for $T>T_{AL}$.

\begin{figure}
\begin{center}
{\parbox[t]{13.5cm}{\epsfxsize 13.5cm
\epsffile{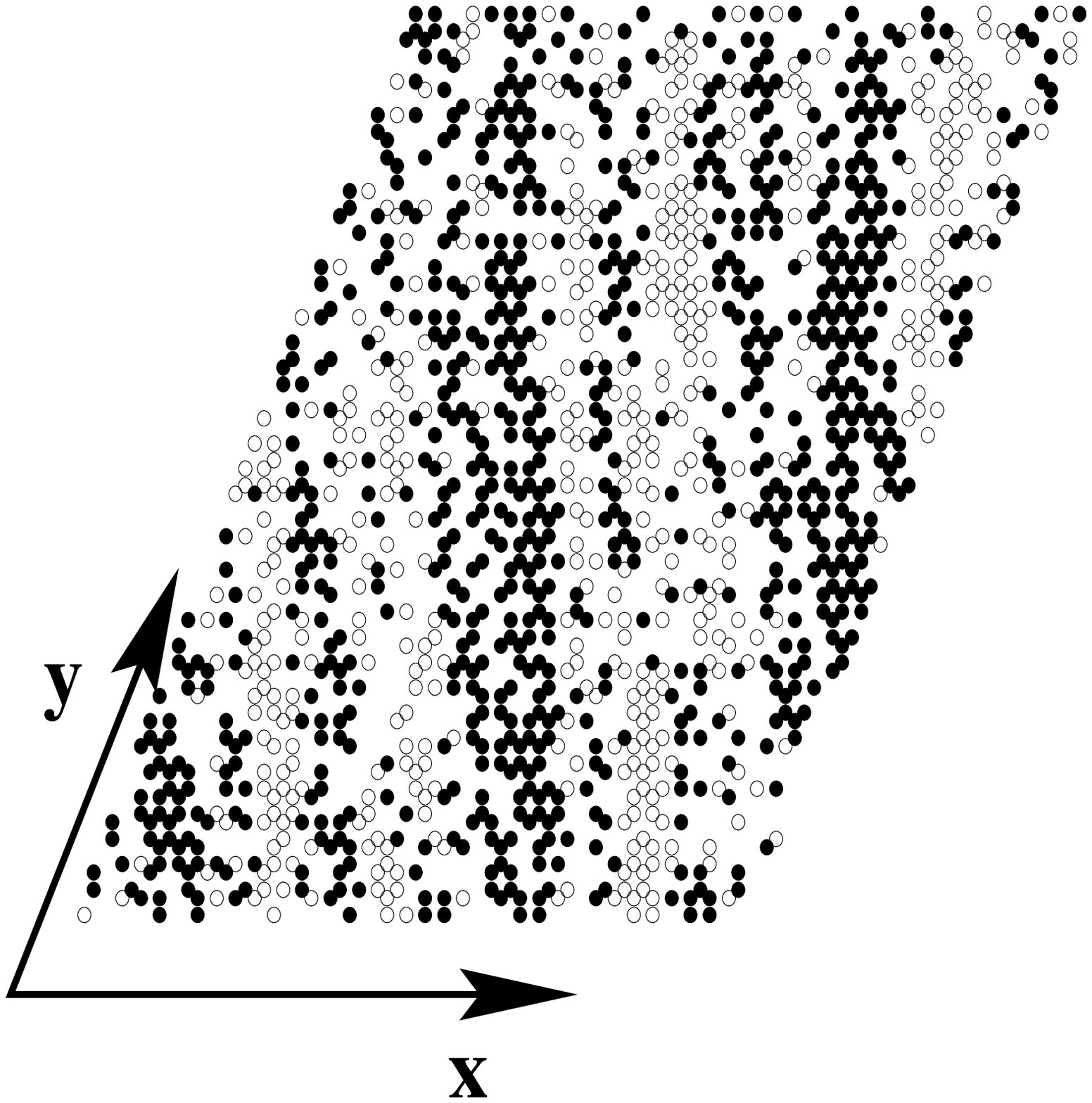}}
}
\end{center}
\protect\caption{
Snapshot of chiralities on each  plaquette of a $36^2$ triangular lattice.
for $T=0.4J$. Filled
circles represent plaquettes with the correct sign, i.e in the same chiral state
 as at $T=0$. Open circles correspond to plaquettes with the wrong sign, that is
 such that the chirality has changed compared to $T=0$. Plaquettes with zero 
chirality  (no symbol) are obtained in-between the two. One clearly sees a stripe 
structure of filled circles and open circles separated by domain walls of zero
chirality. The horizontal ($x$) axis corresponds to the direction of the
$\eta$ bonds.
}
\label{VISURUB}
\end{figure}

The fact that $\gamma^{xx}$ vanishes continuously at $T=T_{AL}$ suggests
that the commensurate-incommensurate transition is not  first
order. The structure of $P(Q_x)$ and the temperature dependence of
$\gamma^{xx}$ for $T\geq T_{AL}$ signal in fact an inhomogeneous
equilibrium state with striped domains in which 
$Q_x=+Q_0(T)$, $Q_x=-Q_0(T)$ or $Q_x=0$\cite{KASH,preprint,UIP1,DW}.
We dub this phase a smectic-like phase\cite{DLK}: in this state, the
elasticity in the $x$ direction vanishes, whereas it is finite in the $y$
direction. The hamiltonian describing long wavelength fluctuations
is smectic-like\cite{DLK}, i.e with a momentum dependence of the 
inverse propagator varying as $Q_x^4$ for the $x$ part and quadratically
with  $Q_y$ for the $y$ part. 
The structure is exhibited in a snapshot of the chiralities of the system
 (Fig (\ref{VISURUB})): chiralities are positive for domains having 
$Q_x=+Q_0(T)$,
negative for domains having $Q_x=-Q_0(T)$ and zero for walls ($Q_x=0$).
 At the transition, for $T=T_{AL}$, the stripe structure is 
stabilized\cite{WS,DN}.

The same characteristics are observed in the entire region ALC of the phase
diagram (see Fig (\ref{PHASEDIAG})).  

So, in contrast to the previous MC study, we find here four phases in the
($\eta , T$) plane. In addition to the (C), (IC) and (P) phases, 
an inhomogeneous thermodynamic state is observed between the (C)
and (IC) phases, 
bounded by lines ALC (see Fig (\ref{PHASEDIAG})). Its presence was
revealed by our method. 

\section{Conclusion}
We have presented a technique  allowing to study numerically the ground
state and the critical regime of finite size inhomogeneous or incommensurate
structures. We illustrate its features on the
commensurate-incommensurate transition of classical 2D XY spins on the
triangular lattice   
\section*{acknowledgments}

Monte Carlo calculations were performed on a Cray C98 thanks to contrat
960162 from IDIS. Support from NATO grant 930988 is acknowledged.

\newpage
\vspace{1truecm}
\newpage

\vspace{-2truecm}
\newpage
\begin{figure}
\begin{center}
{\parbox[t]{5.5cm}{\epsfxsize 5.5cm
}}
\end{center}
\end{figure}

\newpage
\newpage
\newpage
\vspace{-2truecm}
\newpage
\vspace{-4truecm}
\newpage

\end{document}